\documentstyle[revtex]{aps}
\newcommand{\Box}{\hbox{{$\sqcup$}\llap{$\sqcap$}}}

\begin{document}
 \preprint{}
 \draft
\begin{title}
Formation of Solitonic Stars Through Gravitational Cooling
\end{title}

\author{Edward Seidel}

\begin{instit}
National Center for Supercomputing Applications
University of Illinois
Champaign, IL 61820
\end{instit}

and

\author{Wai-Mo Suen}

\begin{instit}
McDonnell Center for the Space Sciences
Washington University
St. Louis, MO 63130
\end{instit}

\receipt{}
\begin{abstract}
We studied the formation of compact bosonic objects through a dissipationless
cooling mechanism.
Implications of the existence of this mechanism are discussed, including the
abundance of bosonic stars in the universe, and the possibility of ruling
out the axion as a dark matter candidate.

\end{abstract}

\narrowtext

I. Introduction.  It has long been known that there exist compact
self-gravitating
soliton-like equilibrium configuration of bosonic fields.$^{1}$
The recent surge of interest in these solitonic objects is
largely due to the suggestion that the dark matter could be
bosonic in nature.  Many particle theories predict that
weakly interacting bosons are abundant in the universe,$^2$ and
may have played a significant role in the
evolution of the universe.

There are two types of self-gravitating compact solitonic objects
made up of bosonic fields known as boson stars$^1$ and oscillating soliton
stars$^3$ (oscillatons).  In many ways, they
are similar to neutron stars.  The simplest example of a boson
star is made up of a complex massive Klein-Gordon scalar field,
with no self-interaction except through gravity:
\begin{equation}
     \Box \phi + m^2\phi = 0 , ~~~~ G_{\mu \nu} = 8\pi \, T_{\mu\nu}(\phi)
\eqnum{1}
\end{equation}
where $T_{\mu\nu}$ is the usual stress energy given by $\phi$ and its
derivatives.  The total mass of a boson star described by
Eq.~(1), ranges from 0 to a maximum of
$M_c=0.633M^2_{\rm Planck}/m$, which is typically smaller than stellar mass.
However, if a quartic self-couple term is included,
even for a small coupling constant,
the mass can be comparable to a neutron star.$^4$

The major difference between a boson star and an oscillaton
is that the latter is made up of a bosonic field without a conserved
current.  The simplest example of an
oscillaton is made up of a real massive Klein-Gordon field, again
described by Eq.~(1), but now with $\phi$ being a real
field.  The non-linearity of these coupled Einstein-Klein-Gordon
equations again gives rise to a non-topological solitonic
solution.  However, unlike the complex field case,
this soliton solution is time dependent, with both the spacetime
geometry $g_{\mu\nu}$ and $\phi$ oscillating in time, in a way
similar to the breather solution to the sine-Gordon
equation.  The existence of a solitonic solution in this
case is particularly interesting, because the axion, being a real
(pseudo) scalar field, is one of the most promising candidates of
dark matter.$^{5}$  When the axion's coupling to other matter fields
and self-couplings can be neglected when the density is low,
it is described by the Eq.~(1).

However, the existence of a solitonic solution is not a guarantee
that such an object can actually be formed in the universe.
It is well-known$^{6}$ that for self-gravitating fields described
by Eq.~(1), there is a gravitational instability analogous to
the Jeans' instability.  However, if there is no
efficient cooling mechanism to get rid of the excess kinetic energy, as
is apparently the case for a system described by Eq.~1,
collapse initiated by this instability
generally only leads to a diffuse viralized ``cloud'',
but not a compact object.
This is an outstanding fundamental question in the study of the bosonic
compact objects.

It is of particular interest to look at the case of axion.
It is generally believed that
axions, having no effective cooling mechanism,
remain in the form of extended gas cloud after separation
from the Hubble flow.$^{7,8}$  Hogan and
Rees$^{8}$ argued that the initial isocurvature
perturbation leads to ``axion miniclusters'' of mass
$\sim 10^{27}$~g, with diameters of order $10^{14}$~cm, which
subsequently undergo classical hierarchical
clustering.

In this paper we show that there {\bf is } a
dissipationless cooling mechanism which very efficiently
leads to the formation of compact bosonic objects.  This
mechanism, which  we call gravitational cooling, is similar to the violent
relaxation of collisionless stellar systems.$^{9}$  A
collisionless stellar system, quite independent of the initial
conditions, would collapse to a centrally dense system by sending
some of the stars to large radius, and
settles into an equilibrium configuration with a more or less
definite distribution.  Likewise, quite independent of the
initial conditions, a scalar field configuration described by
Eq.~(1), will collapse to form a compact soliton star [boson
star for the complex field case, oscillaton for the real
field case], by ejecting part of the scalar field, carrying out the
excess kinetic energy.  In retrospect, it should not be surprising that
there is such a cooling mechanism similar to the violent
relaxation.  The
evolution of a massive scalar field under its self-gravity is in
many ways similar to that of ordinary material bodies,
{\it e.g.,}~in the Jeans' instability analysis.$^{6}$
It has been shown that perturbed boson stars and oscillatons
can evolve back to their
equilibrium configurations by radiating part of the scalar
field.$^{10,3}$  Furthermore, such scalar radiation can drive
the equilibrium configurations on the unstable branch to the
stable branch.$^{10}$

On the other hand, gravitational cooling is different from violent relaxation
in many ways.  For example, gravitational cooling
is (1) efficient even when the ratio of the initial kinetic to potential energy
is
large, (2) much more thorough as a relaxation process, ending on a
unique final state independent of the initial conditions,
and (3) it has a much larger ratio of the initial
to final sizes.  These features will be demonstrated
in the next section, where we present results of a numerical study of
the gravitational cooling mechanism.

The immediate implication of the existence
of the dissipationless gravitational cooling mechanism is that
bosonic solitonic stars can be formed
in our universe.  If the dark matter is described by a classical
bosonic field, it
could be made up of such stars, provided that the localized clouds
separated from the Hubble flow are able to collapse
to objects with masses not larger than a
critical mass $M_c$.

The abundance of such solitonic stars with
astrophysical mass but microscopic size has interesting
consequences on galaxy formation, microwave background power
spectrum, galaxy dynamics, formation of the first stars, among
other things.  A particularly intriguing possibility is that such a
mechanism may rule out the axion as a dark matter candidate.
The axion miniclusters of Hogan and Rees$^{8}$
have a density of $10^{7}$~g/cm$^3$.
At such a density, the
annihilation $(AA \rightarrow \gamma\gamma$), and other dissipative
processes, are not effective,$^{7}$
and the evolution is accurately described by Eq. (1).
Therefore the mini-clusters continue on
collapsing through gravitational cooling after separating out from the Hubble
flow.  For axions with mass
$m\sim 10^{-5}$~eV, the maximum oscillaton mass is
$10^{28}$~g, which is bigger than the total mass of a minicluster.
Hence one might expect that the end point of the collapse is
an oscillaton (but not a black hole).  If there is no further fragmentation
during the collapse of the minicluster, the resulting oscillaton
has a density of $\rho = 10^{24}$~g/cm$^3$ (phase space number
density $n_p \simeq 10^{62}$~cm$^{-3}$).
However, at such high density, the axion is no longer described by a free
Klein-Gordon field.  In particular, we expect the stimulated
decay of axions to be important (single axion decay rate is extremely small,
$ r \sim 10^{-49}~sec^{-1} $ , for
$m_a\sim 10^{-5}$~eV).
The amplification coming from stimulated decay gives a factor of
$\exp (D)$, with$^{12}$
\begin{equation}
  D \sim {\Gamma_\pi\, M_p^2\, V_e \over Rm_\pi^4}
  {f_\pi \over f_a} \eqnum{2}
\end{equation}
where $\Gamma_\pi \sim 8$~eV, $f_\pi \sim 1$~GeV,
$f_a\sim 10^{12}$~GeV (for $m_a\sim 10^{-5}$~eV),
$m_\pi =135$~MeV, $V_e$ is the escape velocity
$(\sim\sqrt{{2GM\over R}})$, and $R$ the radius of the bound
object.  For an oscillaton with the mass of a mini-cluster,
$D\sim 10^{27}$, which simply implies that the collapse driven by
the gravitational cooling ends up in a bright flash.$^{13}$
This suggests that it might not be self-consistent
to have the axions as a dark matter candidate: the axion
has a tendency to form compact objects
(oscillatons) in a short dynamical time scale, but it is unstable in such a
state.

II. Numerical Results.  In this section we present the results of a numerical
study of the
gravitational cooling mechanism for the Klein-Gordon scalar fields,
described by the Eqs. (1)  (see Refs. 3, 10,
for the full set of equations and a description of the numerical methods.)
Sperical symmetry is assumed through out, as our
aim is to demonstrate
the existence of the cooling mechanism but not
the detailed modeling of the formation of a soliton star in the
astrophysical environment.  We do not expect the spherical evolution
to reflect the details of a realistic collapse, as processes like
fragmentation, formation of pancake, {\it etc.} have been
left out.  A generalization of the study to the 3-D case is underway.

In Fig.~1, we show the evolution of a complex
Klein-Gordon scalar field with self-gravitation.  The initial
configuration is taken to be a gaussian distribution
with $\phi=0.0025e^{-r^2/(90)^2}$ and  $\dot{\phi} = 0.9i\phi$
(all lengths are in unit of $1/m$, $m=$~mass of the scalar field).
Darker areas represent higher field strength.  We see that the field collapses
and settles down to a bound state by ejecting part of itself at
each bounce.  The ejected ``scalar radiation'' is shown as the
black strips to the upper right hand corner.
As time goes on, the ejection is less energetic,
and the strips becomes more vertical as they are emitted with a
smaller speed.  By about $t=4000({1\over m})$,
which is about 4~times the free falling time of the initial
configuration,
it settles down into a perturbed boson
star on the stable equilibrium branch$^{10}$ with a total mass of about
$0.56\, M_p^2/m$.  Both the field distribution and the oscillation
frequency match those of a boson star with the same mass studied
in Ref.~10.  The
overall scalar field ejected by the gravitational cooling process
is given by
$(M_{\rm initial} - M_{\rm final})/M_{\rm initial} \simeq 0.13$
We note that the initial mass $M_{\rm initial}$ of this configuration
was $0.644\, M_p^2/m$, which is greater than the maximum mass of
a stable boson star.  Without the mass loss due to scalar radiation
this configuration can only form a black hole.

To show that an initial collapse does not necessarily lead to the
formation of a compact object, we have studied the evolution of
a {\bf massless} scalar field with the same initial condition.
The field collapses, rebounces and completely
disperses to infinity.  No non-singular self-gravitating
solitonic object can be formed with a massless Klein-Gordon
scalar field.$^{15}$  We have also studied the same configuration
for a massive field as above with the self gravitation turned off
({\it i.e.}, $G= 0$ in Eq.~(1) and with a flat space metric), and
the field simply disperses away with no collapse.

The above evolutions are with the fully relativistic systems.  To
better compare the gravitational cooling to violent
relaxation, which is studied usually in Newtonian terms, we take
the weak field post-Newtonian expansion of Eq.~(1).  The
leading order equations in the expansion, describes a
Schr\"odinger type field under
its own self-gravitation.$^{10}$
The kinetic energy $K$ and the potential energy $W$ of the
configuration can be computed,
with the total energy $E = K+W$ a conserved quantity.

We begin with an initial
configuration $\phi = 0.00045 e^{-(r/210)^2}(1+.5sin(\pi/15 r))(1+i)$.
This is spatially much larger than the previous case,
with a strong
spatial perturbation on the initial packet.  In spite of the very
different initial conditions, the
subsequent evolution is similar.  A boson star is
formed with a smaller total mass.  In Fig.~2, we plot the
change of the total energy in time.  The energy is positive and
remains a constant until the scalar radiation comes out from the
boundary of the grid at $r= 1500(1/m)$.  After this time
it settles down to a bound state with negative binding energy.
The ratio of the kinetic energy $K$ to potential energy $W$ is
shown as solid line with a scale given by the right vertical
axis.  We see that initially $K/ |W|$ is larger than 1.  The
ratio decreases as the outward travelling scalar radiation with
large kinetic energy is emitted from the system in the
gravitational cooling process.  It is interesting to compare this
to the violent relaxation, which is inefficient for an initial
value of $K/|W|$ so large, as shown by van Albada$^{16}$ ({\it albeit},
in a 3-D case).  Also in Fig.~2, the total mass of the system
is shown as the dashed line.  The gravitational cooling process emits
about 55\% of the inital mass by the end of the simulation ($t=60000$ ).

The above cases are all with a complex field.  To be relevant to the axion
(and other bosonic dark matter candidates described by real fields,
{\it e.g.,} pseudo Higgs$^{17}$), we show in Fig.~3 the evolution of a real
scalar field with the same
initial data profile as in Fig. 1.  (In this case, as the
field is real, we chose $\dot{\phi} = 0$.)
The gravitational cooling process is the same.
The difference is that the final state has intrinsic
oscillations,$^3$ {\it i.e.,}~it is an oscillaton instead of a
boson star.

III. Discussions and Conclusion.
We demonstrated that there is a dissipation{\bf less} cooling
mechanism which is efficient in forming compact soliton objects
out of bosonic field configurations.  The mechanism is not
specific to a particular kind of field: there is no need to
introduce any particular coupling or damping mechanism.  As long
as the field is massive, its self-gravitational interaction provides the
relaxation towards a bound state in equilibrium.  In the
self-generated time-dependent gravitational potential, energy can
be transferred to a part of the field, ejecting it out of the
system, carrying with it the excess kinetic energy, in a way
similar to the violent relaxation of stellar systems.

The existence of this gravitational cooling mechanism means that
if the dark matter is described by a classical bosonic field,
it is very possible
for it to be in
the form of soliton stars.
In turn it implies that those bosonic
particles that are unstable when the density is high, are likely
to be ruled out as dark matter candidates.

The biggest restriction in our present study is that it is
spherically symmetric, and hence cannot address the question of
fragmentation during collapse, which will be the subject of a
3-D numerical study.  With fragmentation, the gravitational
cooling mechanism should lead to more than one solitonic object with
smaller masses, starting from one collapsing cloud.  These objects
with smaller masses have larger radius and hence lower density.
This leads one to wonder if the axion can be saved by
fragmentation.$^{18}$  We note that for an axionic oscillaton to
non-liminuous, D in Eq. 3 cannot be much larger than 1.$^{12}$
This means that maximum mass of axionic oscillations as dark matter
can only be of order $ 10^{24}g$, having a
radius $10^4$~cm.  Hence in order to avoid being ruled out as
a dark matter candidate, a single minicluster of Hogan and Rees$^{8}$
has to be broken into so many pieces that even after a
cosmological time scale of merging, there are still more than
$10^3$ oscillatons left.  That is not likely to happen
in the generic case.

We would like to thank Ram Cowsik, Leonid Grishchuk, Rocky Kolb,
Martin Rees, Matt Vissor and Clifford
Will for useful discussions.  The work is supported by NSF
Grant No.s ~91-16682.


\pagebreak
\bigskip
\centerline{FIGURES}
\bigskip

\figure{
The evolution of $r^2 \rho$ for a massive, self-gravitating complex scalar
field
is shown.  Due to the self-gravitation, the field collapses
quickly and an perturbed boson star is formed at the center.  The star
oscillates and begins to settle down as scalar material is radiated
through the gravitational cooling process discussed in the text.
\label{fig1}}

\figure{
The evolution of a Newtonian scalar field configuration
is shown.  The ratio $K/ |W|$ of the kinetic to potential energy
is shown as a solid line, and the total mass of the system is shown
as a dashed line.  The inset shows the evolution of the
total energy $E = K + W$.  Initially this configuration is unbound
with positive E, but as part of the field is radiated away the
energy of the remaining configuration becomes negative as a
bound, boson star is formed.
\label{fig2}}

\figure{
The evolution of $r^2 \rho$ for a massive, self-gravitating real
scalar field
is shown.  The evolution is very similar to that shown in Fig. 1 for
a complex configuration with the same initial profile for $\phi$.
\label{fig3}}


\begin{references}
\bibitem{1}
For reviews, see T.~D. Lee, and Y. Pang, Phys. Rep.~{\bf 221},251 (1992),
P. Jetzer, Phys. Rep. {\bf 220 },163 (1992), and
A.~R. Liddle and M.~S. Madsen, Int.\ J.\ Mod.\ Phys.\ D{\bf 1}, 101
(1992).
\bibitem{2}
For review, see J.~R. Primack, D. Seckel, and B. Sadoulet,
Ann.\ Rev.\ Nucl.\ Part.\ Sci.\ {\bf 38}, 751 (1988).
\bibitem{3}
E. Seidel and W.-M. Suen, Phys.\ Rev.\ Lett.\ {\bf 66}, 1659 (1991).
\bibitem{4}
M. Colpi, S.~L. Shapiro, and I. Wasserman,
Phys.\ Rev.\ Lett.\ {\bf 57}, 2485 (1986).
\bibitem{5}
For a review on axion, see, {\it e.g.,} J.~E. Kim,
Phys.\ Rep.\ {\bf 150}, 1 (1987).
\bibitem{6}
M.~Yu. Khlopov, B.~A. Malomed, and Y.~B. Zeldovich,
Mon.\ Not.\ R.\ Astron.\ Soc.\ {\bf 215}, 575 (1985);
M. Bianchi, D. Grasso, and R. Ruffini,
Astron.\ Astrophys.\ {\bf 231}, 301 (1990); D. Grasso, Phys.\ Rev.\ D
{\bf 41}, 29988 (1990);
M.~S. Madsen and A.~R. Liddle,
Phys.\ Lett.\  {\bf B251}, 507 (1991).
\bibitem{7}
M. Fukugita, E. Takasugi, and M. Yoshimura, Z.\ Phys.\ {\bf C27},
373 (1985).
\bibitem{8}
C.~J. Hogan and M.~J. Rees, Phys.\ Lett.\ {\bf B205}, 228 (1988).
\bibitem{9}
D. Lynden-bell, Mon.\ Not.\ R.\ Astron.\ Soc.\ {\bf 136}, 101 (1967).
For recent developments, see D.~N. Spergel and L. Hernquist,
Astrophys.\ J.\ {\bf 397} L75 (1992), and references cited therein.
\bibitem{10}
E. Seidel and W.-M. Suen, Phys.\ Rev.\ D {\bf 42}, 384 (1990).
\bibitem{12}
I.~I. Tkachev, Phys.\ Lett.\ {\bf B191} 41 (1987).
\bibitem{13}
Another process that becomes important at high axion density
comes from the axion's quartic self-coupling, however, it only
enhances the collapse [I.~I. Tkachev, Phys.\ Lett.\ {\bf B261},
289 (1991)] and hence does not affect the above argument.
\bibitem{15}
D. Christodoulou, Commun.\ Math.\ Phys.\ {\bf 105}, 337 (1980),
{\bf 109}, 613 (1987);
D.~S. Goldwirth and T. Piran, Phys.\ Rev.\ D {\bf 36}, 3575 (1987).
\bibitem{16}
van Albada, Mon. Not. R. Astr. Soc. {\bf 201}, 939 (1982).
\bibitem{17}
K. Griest and M. Sher, Phys.\ Rev.\ Lett.\ {\bf 64}, 135 (1990);
Phys.\ Rev.\ D {\bf 42}, 3834 (1990).
\bibitem{18}
Another possibility is that the axion cloud decoupled from the
Hubble flow at an earlier time, so that the axion cluster
has a mass much smaller than that found in Ref. 8.  For
this to be true, there will be
interesting implications on axion phenominology.

\end{references}
\end{document}